\documentclass[twocolumn,printnumbers,amsmath,amssymb,showpacs,10pt]{revtex4}

\newcommand{\be}{\begin{equation}}
\newcommand{\ee}{\end{equation}}
\newcommand{\ba}{\begin{eqnarray}}
\newcommand{\ea}{\end{eqnarray}}
\usepackage{graphicx}
\usepackage{dcolumn}
\usepackage{bm}
\usepackage{setspace}
\usepackage{color}
\begin{document}

\title{Solitons and thermal fluctuations in strongly non-linear solids}
\author{N. Upadhyaya$^{\dag}$, A.M. Turner$^{*}$ and V. Vitelli$^{\dag}$}
\affiliation{$^{\dag}$ Instituut-Lorentz for Theoretical Physics, Universiteit Leiden, 2300 RA Leiden, The Netherlands \\
 $^{*}$Institute for Theoretical Physics, University of Amsterdam,  Science Park 904, P.O. Box 94485, 1090 GL Amsterdam, The Netherlands }

\begin{abstract} 
We study a chain of anharmonic springs with tunable power law interactions as a minimal model to explore the propagation of strongly non-linear solitary wave excitations in a background of thermal fluctuations. By treating the solitary waves as quasi-particles, we derive an effective Langevin equation and obtain their damping rate and thermal diffusion. These analytical findings compare favorably against numerical results from a Langevin dynamic simulation. In our chains composed of two sided non-linear springs, we report the existence of an expansion solitary wave (anti-soliton) in addition to the compressive solitary waves observed for non-cohesive macroscopic particles. \end{abstract}

\maketitle

In linear elastic solids, phonons are the basic mechanical excitations responsible for energy propagation. By contrast, an aggregate of macroscopic grains just in contact with their nearest neighbours constitute a novel elastic material where solitary waves or shocks replace phonons as the basic excitations \cite{Nesterenko_1984,Rosenau_1986,Gomez_2012}. The origin of these strongly non-linear waves can be traced to the fact that, unlike the case of harmonic springs, the repulsive force between two grains in contact does not depend linearly on the relative compression. So far, little effort has been directed to determine the fate of these strongly non-linear excitations in a background of thermal fluctuations because temperature is clearly not a parameter relevant to the elastic response of macroscopic grains. 

However, granular aggregates at zero pressure are just one example of a broader class of materials that can be prepared in a unique mechanical state called \textit{sonic vacuum}  \cite{Nesterenko_1984}. This term originally coined by Nesterenko in the context of strongly non-linear granular chains designate a material characterized by a vanishing elastic moduli and linear speed of sound  \cite{Nesterenko_1984,Nesterenko_Book,Coste2,Daraio2005,Job,Sen}. Grafted colloidal particles \cite{Anand_2007} and ultra-cold atoms in optical lattices \cite{Chang_2013} are microscopic systems that allow for tunable non-linear interactions, while being naturally coupled to a source of fluctuation (thermal or quantum). These fluctuations restore rigidity and generate long wavelength phonon modes \cite{Chaikin_2000,Zhirov_2011}. 
However, the physics of very high amplitude strain propagation is still predominantly non-linear and resembles the state of sonic vacuum perturbed by background fluctuations, even if the interaction potentials  are typically two sided, unlike granular ones. The non-linear regime is particularly relevant for some biological systems, where energy transport occurs through localized non-linear excitations with energy significantly higher than the thermal energy \cite{Davydov_1977,Peyrard_Book}.

Moreover, systems such as polymer networks and colloidal glasses undergoing an unjamming transition are also characterized by vanishing elastic moduli as the coordination number or packing fraction are lowered towards the critical point \cite{OHern}. The effect of thermal fluctuations on the {\it non-linear} response of materials undergoing an unjamming transition is relatively unexplored, despite they are obvious examples of a sonic vacuum state at zero temperature \cite{Mackintosh_2012, berthier,Ning}. Note, that in the case of jamming the linear elastic moduli can be lowered towards zero even if the microscopic interactions are harmonic, simply because there are not enough forces to prevent floppy motions.    

In this article, we focus on strongly non-linear mechanical waves propagating in a background of small thermal fluctuations, a non-equilibrium problem that lies outside the realm of perturbation theory. The starting point of conventional perturbation methods is a linear elastic solid, possibly at finite temperature, perturbed by small anharmonic terms. By contrast, we adopt as a starting point the fully non-linear state of sonic vacuum whose elementary excitations are long-lived solitary waves \cite{remark1}. Subsequently we {\it switch on} temperature as a small perturbation that creates a background of thermal fluctuations. 
 
As a minimal model that is analytically tractable, we study impulse propagation in a one dimensional lattice of non-linear springs with a tune-able power law interaction. By coupling the lattice to a heat bath, we then study the effects of the thermal fluctuations on the leading solitary wave generated in response to an impulse of energy much higher than the background thermal energy. Our approach in a nutshell is to treat the solitary wave as a quasi-particle and derive an effective Langevin equation that describes its stochastic dynamics. We corroborate our analytical predictions for the damping rate and thermal diffusion of the solitary waves with Langevin dynamic 
simulations. The sonic vacuum is usually studied with a chain of non-cohesive beads  that only interact upon compression (one-sided interaction). The system with springs has some of the same properties as the sonic vacuum, since the sonic vacuum is a property of the non-linear power law interaction (without a harmonic term). In addition to the compressive solitary waves seen in a lattice of macroscopic grains with one-sided repulsive interaction, we report an accompanying  anti-solitary wave solution for the lattice of non-linear springs with two sided interactions. 

\begin{figure}
\begin{center}
\includegraphics[width=0.5\textwidth]{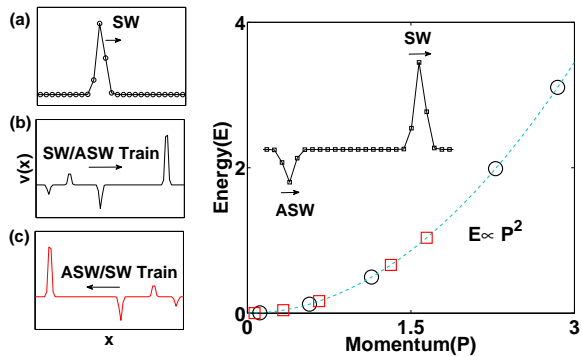}
\caption{\textit{left:}(a) Velocity profile of the compressive  solitary wave (SW) generated in an athermal chain of beads with one sided interaction. (b-c) Velocity profiles showing the formation of a train of SW-ASW pair for two-sided non-linear springs.  A single particle is initially given an impulse to the right, generating a train led by a SW moving in the direction of the impulse (b), while simulatenously generating a symmetric train led by an ASW moving in the opposite direction. \textit{right:} The energy momentum relation for the leading SW/ASW in the three cases shown in the left panel following the energy ($E$) momentum ($P$) relation $E=\frac{P^2}{2 m_{\text{eff}}}$. \textit{inset}: Zoom in of the leading SW-ASW pair from left panel (b).}
\label{ASW}
\end{center}
\end{figure}


\section{The impulse response of non-linear springs}

In Fig.\ (\ref{ASW}), we demonstrate that the compressional solitary wave (SW) excitation discovered by Nesterenko in a chain of non-cohesive beads is also seen in a lattice of springs with two sided interactions. However, unlike the case of a one sided potential, each compressive SW generated in response to an impulse is accompanied by a corresponding expansion solitary wave (formed by local stretching of springs) of the same magnitude but moving in the opposite direction.  This anti solitary wave (ASW) is not sustained by beads interacting with purely repulsive potentials -- the beads would merely loose contact.   

In Fig.\ (\ref{ASW}) left panel, we show the SW/ASW excitations for (a)beads , (b-c) particles connected by springs. In Fig.\ (\ref{ASW}), right panel, we plot, $E$, the {\it total} energy carried by the soliton (after summing over all particles envolved) versus $P$, the {\it total} momentum, for the leading SW/ASW. Note that $E=\frac{P^2}{2 m_{\text{eff}}}$, where the constant $m_{\text{eff}}$ can be viewed as the effective mass of the solitary or antisolitary waves. Inspection of Fig. \ (\ref{ASW}) demonstrates that SW excitations in a lattice of repulsive beads (black circles) have the same effective mass $m_{\text{eff}}$  as a SW and ASW in two-sided springs (red squares). 

As shown in Fig.\ (\ref{ASW}) left panel (b-c), the leading SW-ASW generated in response to an impulse imparted to one of the particles towards the right (direction of arrow) is followed by a train of alternating SW-ASW's excitations, of progressively smaller magnitudes. The smaller SW/ASW's are generated as the particle that initially imparted the impulse, recoils with its left-over energy. This process is repeated several times, leading to the generation of the train of smaller excitations. Since the speed of propagation depends upon the amplitude,  the SW and ASW that start propagating together initially (appearing bounded), eventually separate and become clearly distinguishable.

\section{Langevin Equation}
The classical energy-momentum relation $E=\frac{P^2}{2 m_{\text{eff}}}$ satisfied by the SW motivates the interpretation of the solitary wave as a quasi-particle  \cite{Nesterenko_1984,Job}. For small perturbations, the SW can still be treated as a quasi-particle provided the effects of the perturbations accrue gradually such that the SW retains its functional form. We now apply this adiabatic approximation to derive an effective Langevin equation for the SW quasi-particle when the lattice of springs is coupled to a heat bath. Recall first, the Langevin equation for a particle of mass $m$ undergoing Brownian motion in one dimension is
\begin{eqnarray}
\frac{dx}{dt} &=& v, \nonumber \\
\frac{dE}{dt} &=& -2\frac{\zeta}{m}K + \sqrt{\frac{2\beta^2K}{dt}}N(0,1). \label{Langevin_particle}
\end{eqnarray}
Here, $E,K$ are the total and kinetic energies respectively, $\zeta,\beta$ are the dissipation and diffusion coefficients related via the fluctuation dissipation theorem $\beta^2=2\zeta k_BT/m^2$, where $k_B$ is the Boltzmann constant. $N(0,1)$ is a normal random variable with mean 0 and variance 1,  and encapsulates the effects of random fluctuations  during the time interval $t,t+dt$.  For a free particle of unit mass moving with speed $v$, $E=K=\frac{1}{2}v^2$ and upon substituting in Eq.\ (\ref{Langevin_particle}), we recover the Langevin's equation conventionally expressed as the rate of change of momentum of the particle \cite{Kampen_Book}. 

We now derive an equation analogous to Eq.\ (\ref{Langevin_particle}) for the compressive solitary wave quasi-particle. Let the displacement of a particle from its initial equilibrium position in the continuum limit be $\phi (x,t)$. If we identify the lattice spacing $a$ as a characteristic length scale and $\omega=\sqrt{\frac{k}{m}a^{\alpha-2}}$ as an inverse time scale,  then the equation of motion for the compressive displacement field  $\phi(x,t)$ in dimensionless units  reads
\begin{eqnarray}
\phi _{tt} -\frac{1}{12}\phi _{xxtt} + [{(-\phi _x)}^{\alpha-1}]_{x}= 0 \label{Boussinesq}
\end{eqnarray}
where subscripts denote partial derivatives with respect to space $x$ and time $t$.  Eq.\ (\ref{Boussinesq}) is a simplified form of the Nesterenko equation \cite{Nesterenko_1984,Rosenau_1986}, see appendix A for details. The first two terms express the rate of change of momentum  while the third term represents the force. Although the solitary wave solution to Eq.\ (\ref{Boussinesq}) is not exact (lacking compact support), Eq.\ (\ref{Boussinesq}) provides a good approximation while being analytically more tractable especially since we are interested in keeping the non-linear exponent $\alpha$ general \cite{Rosenau_1986,Gomez_2012}. Note that the equation for the ASW (stretching) is obtained by  modifying the third term $+ [{(-\phi _x)}^{\alpha-1}]_{x} \rightarrow - [{(\phi _x)}^{\alpha-1}]_{x}$ in Eq.\ (\ref{Boussinesq}). 

In analogy with the Langevin equation for a particle, we model the coupling to a heat bath as the sum of two contributions- an external drag and a random fluctuating force, phenomenologically introduced into the equation of motion as :
\begin{eqnarray}
\phi _{tt} -\frac{1}{12}\phi _{xxtt} + [{(-\phi _x)}^{\alpha-1}]_{x}= -\gamma\left(\phi _t - \frac{1}{12}\phi _{txx}\right) + \nonumber \\ \sqrt{\frac{2\gamma}{\alpha\Gamma dt}} \left(\eta(x,t;t+dt) -  \frac{1}{\sqrt{12}}\eta _x(x,t;t+dt)\right) \label{Boussinesq_1}
\end{eqnarray}
where $\gamma=\frac{\zeta}{m\omega}$ is the dimensionless drag coefficient that couples to the momentum $\Pi=\left(\phi _t - \frac{1}{12}\phi _{txx}\right)$. It is useful to define a coupling constant $\Gamma=\frac{ka^{\alpha}}{\alpha k_BT}$ as the ratio of potential to thermal energy in terms of which, the dimensionless diffusion coefficient  is $D=\frac{2\gamma}{\alpha\Gamma}$. The last (noise) term on the right of Eq.\ (\ref{Boussinesq_1}) in conjunction with $\Pi$, satisfies  the fluctuation dissipation theorem \cite{Bishop_2008}.  Here, $\eta (x,t;t+dt)$ is a Gaussian random noise during the time interval $t,t+dt$  with the moments $\langle\eta (x,t;t+dt)\rangle=0$ and $\langle\eta (x,t;t+dt)\eta (x',t';t'+dt')\rangle=\delta(x-x')\delta(t-t')$ respectively,  where angular brackets denotes ensemble averaging. 

To study the propagation of the SW in a background of thermal fluctuations, we now make a working assumption based on the quasi-particle approximation to the SW: whenever the energy of the SW, $E\equiv E_{\text{SW}}\gg k_B T$, the SW functional form is unaltered and only its amplitude $A(t)$ becomes time-dependent. The amplitude $A(t)$ is the collective variable for the SW quasi-particle and other properties of the solitary wave, such as its energy and momentum may be determined from it. Note, the width of the SW is independent of its amplitude and therefore we do not consider its time dependence \cite{Bishop_2008}.

From Eq.\ (\ref{Boussinesq}), the conserved energy is
\begin{eqnarray}
E=  \int dx \ \frac{1}{2}\phi _t^2 + \frac{1}{24}\phi _{tx}^2 + \frac{1}{\alpha}(-{\phi _x})^{\alpha}, \label{Hamiltonian}
\end{eqnarray}
and the energy of the SW may be obtained by integrating Eq.\ (\ref{Hamiltonian}) over the width of the SW of order $W$. (This avoids including the energy of small SW that separate from the main wave). Using Eq. (\ref{Boussinesq}), the rate of change of energy is,
\begin{eqnarray}
\frac{dE}{dt} = \sqrt{\frac{D}{dt}}\int dx\eta(x,t;t+dt)\left(\phi _t+\frac{1}{\sqrt{12}}\phi _{tx}\right) -2\gamma K \nonumber \\ + \sigma(t)  \label{SW_Langevin} 
\end{eqnarray}
where, $K$ is the kinetic part of the energy,
\begin{eqnarray}
K  = \int dx \ \left(\frac{1}{2}\phi _t^2 + \frac{1}{24}\phi _{tx}^2\right). \label{KE}
\end{eqnarray}
The last two terms on the right of Eq.\ (\ref{SW_Langevin}) describe the possible mechanisms of decay of the SW by ``friction'' from the heat bath ($\gamma$) and the "phonon drag" induced by the thermal motion of the chain $\sigma(t)$. The first term is the fluctuating part of the energy. In the following, we make the assumption (verified numerically) that the coupling to the heat bath is more important and therefore, ignore $\sigma(t)$ (see SI for details).

Solving for the SW solution from Eq.\ (\ref{Boussinesq}), we find $\phi _t = A\psi$, where 
\begin{eqnarray}
\psi(x,t)=\text{sech}^{\frac{2}{\alpha-2}}\left(\frac{x-V_st}{W}\right) \label{SW_sol}
\end{eqnarray}
 is the functional form of the SW with amplitude $A$, speed $V_s=A^{\frac{\alpha-2}{\alpha}}$ and width $W=\frac{1}{\sqrt{3}(\alpha-2)}$ in units of the lattice spacing, see appendix A, subsection 1, for details. The SW energy, kinetic energy, and momentum may now be expressed in terms of the collective variable $A$:
\begin{eqnarray}
E=A^2\int dx\ \psi^2(x,t)=A^2I_E \label{SW_E}
\end{eqnarray}
and from the the virial theorem,  
\begin{eqnarray}
K=\frac{\alpha}{\alpha+2}E=\frac{\alpha}{\alpha+2}I_E A^2
\end{eqnarray}
Additionally, the solitary wave momentum is 
\begin{eqnarray}
P=A\int dx\ \psi(x,t)=AI_P.
\end{eqnarray}
Here, 
\begin{eqnarray}
I_E = \int dx\ \text{sech}^{\frac{4}{\alpha-2}}\left(\frac{x}{W}\right) \\
I_P = \int dx\ \text{sech}^{\frac{2}{\alpha-2}}\left(\frac{x}{W}\right),
\end{eqnarray}
 are constants obtained by integrating over all space \cite{Rosenau_1986}. 

Substituting for $E$ and $K$ in terms of $A$, we cast Eq.\ (\ref{SW_Langevin})  into the form of an ordinary Langevin equation (with additive noise) for the collective variable $A(t)$,
\begin{eqnarray}
\frac{dA}{dt} =\sqrt{\frac{2\gamma}{\alpha\Gamma I^2_E A(t)^2 dt}}\int dx\ \eta(x,t;t+dt)\left(\phi _t+\frac{1}{\sqrt{12}}\phi  _{tx}\right)  \nonumber \\ -\frac{\alpha\gamma}{\alpha+2}A \label{Soliton_Langevin}
\end{eqnarray}
where, $\phi\equiv\phi (x,t)$. Eq.\ (\ref{Soliton_Langevin}) is the central result of our work whose analytical predictions we derive and test numerically in the next sections. The first term can be written as $\frac{1}{\sqrt{dt}}\eta_A(t, t+dt)$, where $\eta_A$
is a white noise signal; that is, its correlations are given by $\langle\eta_A(t)\eta_A(t')\rangle=\frac{2\gamma}{(\alpha+2)I_E\Gamma}\delta(t-t')$.  Using
the fact that the correlations of $\eta(x,t)$ are described by delta-functions,
the correlations of $\eta_A(t)$ can be related to the kinetic energy Eq. (\ref{KE}), which can be replaced by $\frac{\alpha}{\alpha+2}I_EA(t)^2$. 

\section{Time dependence of mean and variance}

Taking the expectation value (ensemble average) of Eq.\ (\ref{Soliton_Langevin}), we find 
\begin{eqnarray}
\frac{d\langle A\rangle}{dt} = -\frac{\alpha\gamma}{\alpha+2}\langle A\rangle, \label{Soliton_mean}
\end{eqnarray}
where, owing to the noise term $\eta(x,t;t+dt)$ (which acts between times $t;t+dt$) and $\phi _t(x,t)$ (which is a solution at time $t$) being statistically independent, the expectation value $\langle\eta(x,t;t+dt)\phi _t(x,t)\rangle=0$. Consequently, the solitary wave amplitude decays as 
\begin{eqnarray}
\langle A\rangle = A_{0}e^{-\frac{\alpha\gamma}{\alpha + 2}t} \label{SW_damping}
\end{eqnarray}
where, $A_{0}$ is the initial solitary wave amplitude. Note, the effective damping rate $\gamma'=-\frac{\alpha\gamma}{\alpha + 2}$ is independent of inverse temperature $\Gamma$ but rescales with the exponent of the non-linear potential $\alpha$. 

Similarly,  we solve for the variance of the solitary wave amplitude or equivalently, the variance in the square root of energy. 
Re-defining, $D=\frac{\gamma}{2\alpha\Gamma I^2_E}$, we solve for the variance in the solitary wave amplitude by first evaluating the differential $d[A^2]=A^2(t+dt)-A^2(t)$, by substituting $A(t+dt) = A(1-\frac{\alpha\gamma}{\alpha + 2} dt)+\sqrt{Ddt}\int dx\ \eta(x,t;t+dt)\left(\psi+\frac{1}{\sqrt{12}}\psi _{x}\right)$ from Eq.\ (7) and retaining terms to order $0(\sqrt{dt})$\cite{Gillespie_Book,Lemons_Book}, 

\begin{widetext}
\begin{eqnarray}
d[A^2] =  -\frac{2\alpha\gamma}{\alpha+2}A^2dt + 2A\sqrt{Ddt}\int dx\ \eta(x,t;t+dt)\left(\psi+\frac{1}{\sqrt{12}}\psi _(x)\right)+ \nonumber \\ Ddt\int\int dx dx' \left(\psi(x,t)+\frac{1}{\sqrt{12}}\psi _{x}(x,t)\right)\left(\psi(x',t)+\frac{1}{\sqrt{12}}\psi _{x'}(x',t)\right).
\end{eqnarray}
\end{widetext}
Taking the expectation value, the second term on the right vanishes (as discussed for the mean) and using the property that the noise term is delta-correlated in space, we obtain
\begin{eqnarray*}
d[\langle A^2\rangle] =  -\frac{2\alpha\gamma}{\alpha+2}\langle A^2\rangle dt +  Ddt\int dx \ \left(\psi+\frac{1}{\sqrt{12}}\psi _{x}\right)^2.
\end{eqnarray*}
The last term when expanded gives twice the solitary wave kinetic energy $2K$, see Eq.\ (\ref{KE}),  plus an integral  $\frac{2}{\sqrt{12}}\int dx\psi\psi _x $, that vanishes by symmetry for the SW solution.  Moreover, the SW kinetic energy is related to its total energy via the virial relation $K=\frac{\alpha}{\alpha+2}E$. Hence, we obtain the ordinary differential equation correct to order $dt$ -
\begin{eqnarray}
\frac{d\langle A^2\rangle}{dt} = -\frac{2\alpha\gamma}{\alpha+2}\langle A^2\rangle +  2DI_E\frac{\alpha}{\alpha+2}.
\end{eqnarray}
Solving, the differential equation subject to the initial condition $\langle A^2\rangle_{t=0}=A^2_{0}$ and substituting for $D$, we obtain,
\begin{eqnarray}
\langle A^2\rangle = A^2_{0}e^{-\frac{2\alpha\gamma}{\alpha+2} t} + \frac{1}{2I_E\alpha\Gamma}\left(1-e^{-\frac{2\alpha\gamma}{\alpha+2} t}\right).
\end{eqnarray}
Using Eq. \ (\ref{SW_damping}),  this may be expressed as 
\begin{eqnarray}
\text{var}(A)=\langle A^2\rangle - \langle A\rangle^2=  \frac{1}{2I_E\alpha\Gamma}\left(1-e^{-\frac{2\alpha\gamma}{\alpha+2} t}\right).
\end{eqnarray}
Using the relation in Eq.\ ( \ref{SW_E}), we rewrite the above equation as 
\begin{eqnarray}
\text{var}(\sqrt{E}) =  \frac{1}{2\alpha\Gamma}\left(1-e^{-\frac{2\alpha\gamma}{\alpha+2} t}\right) \label{Var}.
\end{eqnarray}

The coefficient $\frac{1}{2\alpha\Gamma}$ reduces to $\frac{k_BT}{2}$ when the energy is not measured in units of $ka^\alpha$, so this expression is analogous to the velocity variance of a Brownian particle. Note, for large $\alpha$, the SW is effectively one particle wide and thus Eq.\ (\ref{Var}) captures the correct thermal equilibration of the particle energy with the heat bath.  
However, for the dynamics of the SW, Eq.\ (\ref{Var}) is only useful as long as the SW is identifiable against the background thermal energy, that is $E_{\text{SW}}\gg\Gamma^{-1}$. 

\begin{figure}
\begin{center}
\includegraphics[width=0.5\textwidth]{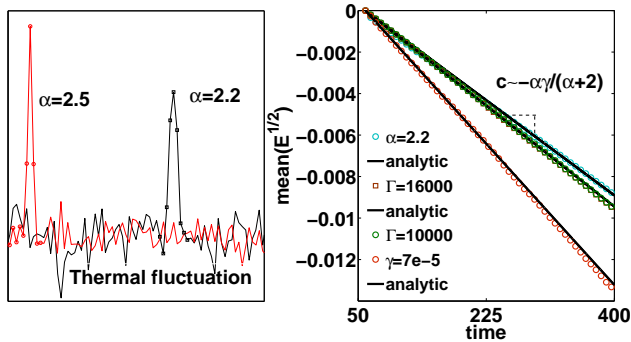}
\caption{\textit{left}: Snapshot of two solitary waves in a background of thermal fluctuations for $\alpha=2.5$ (red) and $\alpha=2.2$ (black). \textit{right}:  the attenuation of the solitary wave as a function of time for various values of $\gamma$,$\Gamma$, and $\alpha$. When not indicated, $\alpha=2.5,\Gamma=10^4$ and $\gamma=5\times 10^-5$.
The data for these values is shown with green circles; the other data corresponds to changing one parameter, $\alpha=2.2$ (black circles),  $\Gamma=1.6\text{x}10^{4}$, (brown squares) and $\gamma=7\text{x}10^{-5}$ (red circles) compared with the analytic expression in Eq.\ (\ref{SW_damping}) (log scale) represented by solid lines. The initial $E_{\text{SW}}=0.5$. }
\label{damping}
\end{center}
\end{figure}


\section{Simulations}

\begin{figure}
\begin{center}
\includegraphics[width=0.5\textwidth]{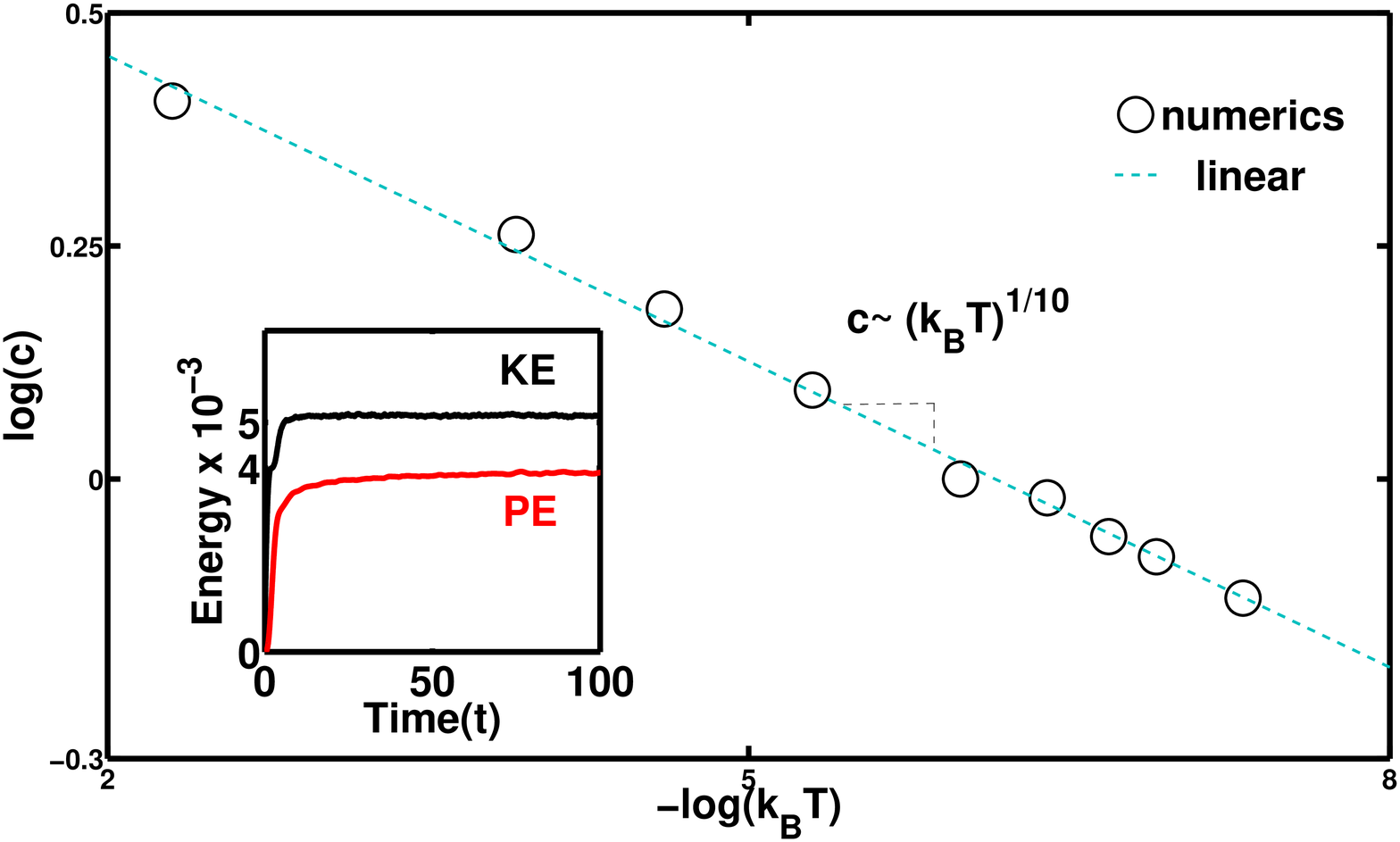}
\caption{The sound speed computed from the dispersion curves for a range of $\Gamma$ (inverse temperature) for $\alpha=\frac{5}{2}$. The circles are from numerical simulation while the dashed blue line is a linear fit, giving a slope of 0.11 for $\alpha=\frac{5}{2}$,  close to the expected value $\frac{\alpha-2}{2\alpha}$. The inset shows the kinetic and potential energy approaching thermal equilibrium, where their ratio satisfies the virial relation $\text{KE}/\text{PE}=\alpha/2$.}
\label{dispersion}
\end{center}
\end{figure}

We consider a one dimensional chain consisting of $N = 1024$ particles each having a mass $m$ placed regularly on a lattice with spacing $a$ (spring rest length) interacting pair-wise with a nearest neighbour interaction $V(\delta) = \frac{K}{\alpha}\left(\delta\right)^{\alpha}$, where $\delta$ is the compression/stretching induced during the dynamics.  We model the coupling to the heat bath by numerically integrating Eq.\ ( \ref{Langevin_particle}) for each particle using the velocity-verlet algorithm \cite{Allen_Book}.  In thermal equilibrium the mean kinetic energy is KE$\sim \frac{k_{B}T}{2}$ and potential energy is PE$\sim \frac{k_{B}T}{\alpha}$, where their ratio satisfies the virial relation, see  Fig.\ (\ref{dispersion}), inset for $\alpha=5/2$. In the following, all numerical data is presented in dimensionless units, ensemble averaged over 1000 samples.

\subsection{Fluctuation induced rigidity}

To extract the equilibrium properties in the thermalized state, we  define the longitudinal current density  of particles as $j(x,t) = \frac{1}{\sqrt{N}}\sum^{N}_{i=1}v_i(t)\delta(x-x_i(t))$, and its Fourier transform $j(k,t) = \frac{1}{\sqrt{N}}\sum^{N}_{i=1}v_{i}(t)e^{ikx}$, where $k$ is  the longitudinal collective mode along the $x$-direction.  Thus, the corresponding longitudinal  current density auto-correlation functions is $C(k,t)=\langle j^*(k,0)j(k,t)\rangle, \label{def-correlation} $ where the angular brackets denote ensemble averaging over the initial time. The longitudinal power spectral density is then obtained as the Fourier transform of the respective current density auto-correlation functions as, $P(k,\omega)=\int^{\infty}_{-\infty}dt\ e^{i\omega t}C(k,t)$. The Fourier transforms defined above are evaluated using fast Fourier transform from simulation data. The sound speeds in Fig.\ (\ref{dispersion}) correspond to the linear part of the dispersion curves, obtained by projecting the power spectral densities on the frequency ($\omega$)- wavenumber ($k$) plane. 

In Fig.\  \ref{dispersion}, we plot the sound speed from the slope of the dispersion curves for $\alpha=5/2$ for a range of $\Gamma$. At thermal equilibrium, the mean kinetic energy and hence the temperature $T$ satisfy the virial relation $T \sim \delta^{\alpha}_T$, where $\delta _T$ is the average displacement of the particles induced by thermal fluctuations. Defining the sound speed $c$ as the second derivative of the induced potential energy leads to the relation, $c^2 \sim T^{\frac{\alpha-2}{\alpha}}$ \cite{Zhirov_2011}. For $\alpha=\frac{5}{2}$, we find $c\sim\Gamma^{\frac{-1}{10}}\sim(k_BT)^{1/10}$, closely matching the linear fit in SI Fig.\ ( 1). Thus, coupling the lattice of non-linear springs that is initially in its state of sonic vacuum (implying the absence of linear sound) to a heat bath, leads to hydrodynamical sound modes with a linear sound speed that scales with the temperature of the heat bath \cite{Zhirov_2011}. Note, setting $\alpha=2$ (harmonic springs) yields a sound speed that is independent of temperature while the limit $\alpha\rightarrow \infty$ yields $c\sim (k_BT)^{1/2}$, a result in agreement with the entropic elasticity for hard sphere colloidal crystals \cite{Chaikin_2000}.
\begin{figure}
\begin{center}
\includegraphics[width=0.5\textwidth]{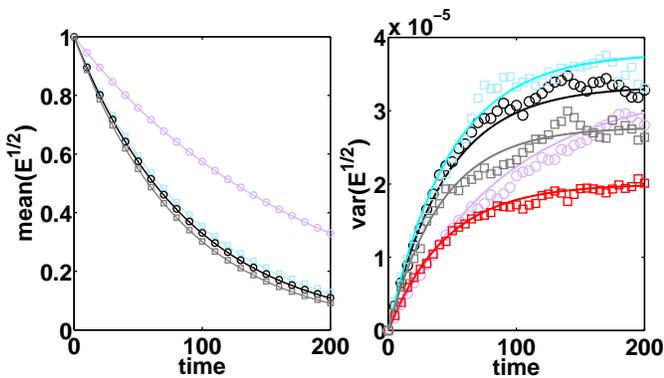}
\caption{\textit{left}: the numerically obtained mean solitary wave energy for $\alpha=2.5,\gamma=0.01$ (purple circles), $\alpha=2.2,\gamma=0.02$ (blue squares), $\alpha=2.5,\gamma=0.02$ (black circles), $\alpha=3.0,\gamma=0.02$ (gray squares) decaying exponentially compared against the analytical expression (solid curves). The mean decay rate is independent of the temperature $\Gamma^{-1}$. 
 \textit{right}:  the numerically obtained varaince of the solitary wave energy for $\alpha=2.2,\gamma=0.02,\Gamma=6000$ (blue squares), $\alpha=2.5,\gamma=0.02,\Gamma=6000$ (black circles), $\alpha=3.0,\gamma=0.02,\Gamma=6000$ (grey squares), $\alpha=2.5,\gamma=0.01,\Gamma=6000$ (purple circles) and $\alpha=2.5,\gamma=0.02,\Gamma=10000$ (red squares)  compared against the analytical expression Eq.\ \ref{Var} (solid curves).}
\label{variance}
\end{center}
\end{figure}

\subsection{Comparison with analytics}

Once the lattice reaches thermal equilibrium, we excite a solitary wave (SW) by imparting one of the particles an initial energy of order $E_{\text{SW}}=0.5$ in dimensionless units. In Fig.\ (\ref{damping})  left panel, we show a snapshot of two SWs at the same time,  propagating in a background of thermal fluctuations for $\alpha=2.5$ (red) and $\alpha=2.2$ (black). We see that the SW with lower $\alpha$ is wider  and moves faster for the given amplitude, in qualitative agreement with the analytic widths $W\sim\frac{1}{\sqrt{3}(\alpha-2)}$ and speeds $V_s\sim A^{\frac{\alpha-2}{\alpha}}$.  

In Fig.\ (\ref{variance}) left panel, we plot the numerical data (symbols) for the attenuation of the SW amplitude as a function of time for various values of $\gamma$ and $\alpha$ and we find a very good match to the analytic expression in Eq.\ (\ref{SW_damping}) (solid curves). For the range of $\Gamma$ explored, we find the damping rate is independent of temperature ($\Gamma$) but depends on the environmental drag $\gamma$ and $\alpha$.

In Fig.\ (\ref{variance}) right panel, we show the increase in the variance of SW amplitude (or the square root of its energy) as a function of time for multiple values of $\alpha$ , $\gamma$ and $\Gamma$ obtained numerically (symbols) and compare them with the complete analytical solution  Eq.(\ref{Var}) finding good agreement. Notice, the final value of the variance correctly approaches the thermal energy, as expected for a Brownian particle. However, since the solitary wave is a dynamical object that decays under the influence of the external drag, once the solitary wave energy becomes comparable to the background thermal energy, it is no longer meaningful to consider it as a Brownian particle.

\appendix

\section{Continuum approximation}

In this section, we will review the Rosenau approximation to  the Nesterenko solitary wave solution, that is valid for any general non-linear potential \cite{Rosenau_1986,Gomez_2012}.

Here, we adopt as our starting point the Lagrangian for a one dimensional chain of identical spheres that are just touching each other, i.e., in the limit $\delta _0\rightarrow 0$ -
\begin{eqnarray}
L = \sum _n 1/2m\dot{u}^2_n - K/\alpha\left(\frac{u_n - u_{n+1}}{a}\right)^{\alpha} \label{A_Lagrangian}
\end{eqnarray}
where, $u_n$ is the displacement of the $n-$th sphere from its equilibrium position, $a=2R$ is the equilibirum lattice spacing and $K$ is the spring constant. In order to avoid doing a Binomial expansion in powers of $\alpha$, we will define the continuum field variable as
\begin{eqnarray}
a\phi'(n+1/2)=u_{n+1} - u_{n},
\end{eqnarray}
where, primes denote derivative with respect to $x$. We now take the continuum limit, i.e., $u_n\rightarrow u(x)\equiv u$ and Taylor expand the right hand side about $x+a/2$:
\begin{eqnarray}
 a\phi'(x) \approx u+ a/2u' + a^2/8u'' + a^3/48u''' - u  + a/2u' - \nonumber \\  a^2/8u'' +  a^3/48u'''.
\end{eqnarray}
Integrating both sides once with respect to $x$, we obtain
\begin{eqnarray}
\phi(x) &=& u + a^2/24u'',\\
            &=& \left(1+ a^2/24\frac{d^2}{dx^2}\right)u(x).
\end{eqnarray}
Inverting the differential operator, we obtain
\begin{eqnarray}
u(x) \approx \phi - a^2/24\phi''.
\end{eqnarray}
Thus, in the contiuum limit, the Lagrangian becomes 
\begin{eqnarray}
L/m &=& \int \ dx 1/2\dot{u}^2(x) - K/(m\alpha)\left(\phi'(x)\right)^{\alpha} \\
       &=&  \int \ dx  1/2\dot{\phi}^2 -a^2/24\dot{\phi}\dot{\phi''} - K/(m\alpha)\left(\phi'\right)^{\alpha}. \label{Lagrangian}
\end{eqnarray}
By using the Euler-Lagrange equation, we obtain the equation of motion as
\begin{eqnarray}
\ddot{\phi} - a^2/12\ddot{\phi''} + \frac{K}{m}\left[(-\phi')^{\alpha-1}\right]'=0.
\end{eqnarray}
Note, $\phi$ here corresponds to the continuum displacement field. The corresponding equation in the strain field $\delta=-\phi'$ reads
\begin{eqnarray}
\ddot{\delta} - a^2/12\ddot{\phi''} - \frac{K}{m}\left[\delta^{\alpha-1}\right]''=0.\label{Ross}
\end{eqnarray}
 Upon substituting $\delta=-\phi _x$ for the compressive SW or $\delta=\phi _x$ for the expansive ASW, we  find the same functional forms for the solitary wave solutions in both cases. Here, $\delta (x,t)$ represents the compression of two adjacent particles i.e., the strain field. 

\subsection{Solitary wave solution}

The solitary wave solution of Eq.\ \ref{Ross} can be obtained by looking for propagating solutions of the form $\delta(x,t)=\delta(x-V_st)$:
\begin{eqnarray}
\frac{V_s^2a^2}{12}\delta'' - V_s^2\delta + \frac{K}{m}\delta^{\alpha-1}=0,
\end{eqnarray}
which can be expressed in the form of Newtons's-like equation
\begin{eqnarray}
\delta'' = -\frac{12}{V_s^2a^2}\left[-V_s^2\delta + \frac{K}{m}\delta^{\alpha-1}\right] = -\frac{dW}{d\delta}.
\end{eqnarray}
Multiplying both sides by $\delta'$ and integrating, 
\begin{eqnarray}
\int dx \delta'\delta'' &=&  \int dx -\frac{dW}{d\delta}\delta', \\
\int \frac{1}{2}d(\delta')^2 &=&  \int -dW,\\
\frac{1}{2}(\delta')^2 &=&  -W(\delta).
\end{eqnarray}
Integrating again, we find 
\begin{eqnarray}
\int \frac{d\delta}{\sqrt{-2W}} = \int dx.
\end{eqnarray}
Substituting, $W(\delta) = \frac{12}{V^2_sa^2}\left[\frac{K}{m\alpha}\delta^{\alpha} - \frac{V^2_s}{2}\delta^2\right]$, and writing for brevity $A=\frac{12}{V^2_sa^2}\frac{K}{m\alpha}$ and $B=6/a^2$, we need to integrate
\begin{eqnarray}
\int \frac{d\delta}{\sqrt{2}\delta\sqrt{B-A\delta^{\alpha-2}}} = x.
\end{eqnarray}
Making the change of variables, $z^2 = B - A\delta^{\alpha-2}$, we find
\begin{eqnarray}
\frac{-\sqrt{2}}{(\alpha-2)}\int \frac{dz}{(\sqrt{B})^2-z^2} &=& x. 
\end{eqnarray}
Therefore,
\begin{eqnarray}
x = \frac{-1}{\sqrt{2B}(\alpha-2)}\left[\int \frac{dz}{\sqrt{B} - z} + \int \frac{dz}{\sqrt{B} + z}\right], 
\end{eqnarray}
that yields,
\begin{eqnarray}
s = -\sqrt{2B}(\alpha-2)x = \ln\frac{\sqrt{B}+z}{\sqrt{B}-z}, 
\end{eqnarray}
or
\begin{eqnarray}
z = -\sqrt{B}\frac{1-\exp(-s)}{1+\exp(-s)}.
\end{eqnarray}
Squaring and substituting $z^2= B - A\delta^{\alpha-2}$,
\begin{eqnarray}
B-A\delta^{\alpha-2} = B\frac{\exp{s/2}-\exp{-s/2}}{\exp{s/2}+\exp{-s/2}},
\end{eqnarray}
that yields,
\begin{eqnarray}
\delta^{\alpha-2} = \frac{B}{A}\text{sech}^2(s/2).
\end{eqnarray}
Therefore, the solitary wave solution is 
\begin{eqnarray}
\delta = \left(\frac{m\alpha V^2_s}{2K}\right)^{\frac{1}{\alpha-2}}\text{sech}^{\frac{2}{\alpha-2}}\left(\frac{\sqrt{3}}{2a}(x-V_st)\right).\label{Ross_sol}
\end{eqnarray}

\section{Solution from discrete equations of motion}

The energy and the fluctuations of the solitary wave can also be derived for a discrete chain, without making the continuum approximation.
From Eq.\ (\ref{A_Lagrangian}), the equation of motion is 
\begin{equation}
m\ddot{u}_n=K(u_{n-1}-u_n)^{\alpha-1}-K(u_{n}-u_{n+1})^{\alpha-1}. \label{discrete_EOM}
\end{equation}
A solitary wave solution to the above equation of motion has the form of a wave moving at a constant speed $V_s$:
\begin{equation}
u_n(t)=\frac{A}{V_s}f(na-V_st). \label{eq:train}
\end{equation}
where $A$ is the amplitude of the SW and $f$ is a function that describes the shape of the SW (Since we define the amplitude as the maximum speed of the particles enveloped by the solitary wave, the displacement $u_n(t)$ will be proportial to $A/V_s$.) 

%

In analogy with Eq.\ (\ref{Langevin_particle}), we now couple the discrete equation of motion Eq.\ (\ref{discrete_EOM}) to a source of Gaussian noise and drag :
\begin{eqnarray}
m\ddot{u}_n=K(u_{n-1}-u_n)^{\alpha-1}-K(u_{n}-u_{n+1})^{\alpha-1}  -\gamma \dot{u}_n \nonumber  \\ +\sqrt{2\gamma k_BT}\eta_n(t). \label{discrete_Noise}
\end{eqnarray}
Here, the noise satisfies $\langle \eta_n(t)\eta_n(t')\rangle=\delta(t-t')$. The coefficient of the noise ensures that in equilibrium the particles satisfy the fluctuation-dissipation theorem such that their average kinetic energy is $\frac{1}{2}k_BT$. 

From Eq.\ (\ref{A_Lagrangian}),  the energy of the chain is $E=\sum_n \frac{1}{2}m\dot{u}_n^2+\frac{K}{\alpha}(u_n-u_{n+1})^\alpha$. If we restrict the summation to particles enveloped by the solitary wave such that the sum does not include particles that are far away from the SW, then $E$ corresponds to the energy of the SW. Multiplying Eq.\ (\ref{discrete_Noise}) by $\dot{u}_n$, we obtain the rate of change of energy
\begin{equation}
\frac{dE}{dt}=\sum_n -\gamma \dot{u}_n^2+\sqrt{2\gamma k_BT}\eta_n(t)\dot{u}_n,
\label{eq:energyconservation}
\end{equation}
where, the contribution of terms of the form $(u_{n\pm 1}-u_n)^{\alpha-1}\dot{u}_{n\pm 1}$ (i.e., with index $n-1$ and $n+1$) cancel out upon summation over $n$. 

Consistent with the quasi-particle interpretation of the SW, we again  assume that the SW in a background of noise and drag,  is still given approximately by Eq. (\ref{eq:train}). Therefore, we write the terms in Eq.\ (\ref{eq:energyconservation}) in terms of the SW amplitude. The SW energy is $E=I_E A^2$, and the first term on the right hand side is proportional to the kinetic energy, which is  $2\frac{\alpha}{(\alpha+2)}E$ by the virial theorem. Thus the energy decays at a rate $\frac{\alpha}{\alpha+2}$ times the decay rate for
the velocity of a single particle ($\frac{\gamma}{m}$).  The SW amplitude therefore decays as:
\begin{equation}
\dot{A}=-\frac{\gamma\alpha}{m(\alpha+2)}{A}- \sqrt{\frac{\gamma k_BT}{2I_E^2}}\sum_n\eta_n(t)f'(na-V_s t).
\end{equation}
A term that compensates for the diffusion of $A$ is omitted here (this term is derived in Ito calculus), but it is only
a small amount that is negligible if $I_E A^2\gg k_BT$. Now, the different contributions in the last term just add up to a new noise term $\lambda\eta(t)= \sqrt{\frac{\gamma T}{2I_E^2}}\sum_n\eta_n(t)f´(na-V_s t)$, which is also not correlated in time. (Assume that $\lambda$ is the amplitude of the noise while $\eta$ has a noise with one unit of strength.)  The original noise term $\sum \eta_n(t) \dot{x}_n$ has correlations because $x_n$ depends on the noise at an earlier time.  But if the noise mostly just changes the amplitude of the SW
by a random amount, and does not cause the particles in it to be displaced randomly relative to each other, then these correlations are not too important, and that is why they cancel out in the equation for $A$.
The variance of the noise is therefore
\begin{equation}
\lambda^2\langle \eta(t)\eta(t´)\rangle=\frac{\gamma k_BT}{2 I_E^2}f´'(na-V_s t)^2\delta(t-t')
\end{equation}
so $\lambda^2=\frac{\gamma T}{2I_E^2}f'(na-V_s t)^2$.  This can be determined by the virial theorem since it is proportional to the kinetic energy of the SW, hence $\lambda^2=\frac{\gamma k_BT}{I_E m}\frac{\alpha}{\alpha+2}$. 
Therefore the amplitude of the soliton satisfies
\begin{equation}
\dot{A}=-\frac{\gamma \alpha}{m(\alpha+2)}A+\sqrt{\frac{\alpha}{\alpha+2}\frac{\gamma T}{I_E m}}\eta(t).\label{discrete_A}
\end{equation}
Upon taking the mean and variance of the amplitude from Eq.\ (\ref{discrete_A}), we recover the solutions derived using continuum approximation. 

\textbf{Acknowledgments} NU acknowledges financial support from FOM.


\begin{thebibliography}{99}
\bibitem{remark1} We are using ``soliton" and ``solitary wave" interchangeably to refer to a localized lump of of energy; we do not assume anything about the equations being integrable.
\bibitem{remark2} For many waves or solitons, differentiating $E$ with respect to $P$ would give the group velocity or the speed of the wave's propagation, but this does not work for this case.  ($P$ is not the canonical momentum.)
\end{thebibliography}
\end{document}